\documentclass[12pt,a4paper]{article}
\usepackage{subcaption}
\usepackage{ragged2e}
\usepackage{mathtools}
\usepackage{physics}
\usepackage{fullpage}
\usepackage{graphicx}
\usepackage{epstopdf}
\usepackage{amssymb}
\newtheorem{definition}{Definition}

\newtheorem{theorem}{Theorem}
\newtheorem{proof}{Proof}
\usepackage{lipsum}
\usepackage{changepage}
\usepackage{amsmath,lipsum}
\usepackage{geometry}
\usepackage{setspace}
\usepackage{todonotes}
\usepackage{kantlipsum}

\allowdisplaybreaks

\begin{document}
\begin{flushleft}

\LARGE
\begin{center}
\textbf{Should monogamy of entanglement be defined via equalities or inequalities?} 
\end{center}

\normalsize
\textit{Alexey Lopukhin}

\footnotesize
\textit{School of Mathematical Sciences, University of Nottingham}

\begin{center}
\LARGE
Abstract.
\end{center} 

\begin{adjustwidth*}{4em}{4em}
\normalsize
\justify
This work focuses on the entanglement quantification. Specifically, we will go over the properties of entanglement that should be satisfied by a "good" entanglement measure. We will have a look at some of the propositions of the entanglement measures that have been made over the years. Then we will be ready to discuss the proposals of the mathematical representations of another property of entanglement, called monogamy. We will introduce some definitions of monogamous entanglement measures that were proposed and compare them. As an original observation of mine (see page 15, Proof 1), I will also prove that the inequalities (23) and (24) from [C. Lancien, S. Di Martino, M. Huber, M. Piani, G. Adesso and A.Winter Phys. Rev. Lett., 117:060501 (2016).] automatically show that the entanglement of formation and the regularised entropy of entanglement are monogamous entanglement measures in the sense of the definition that was given in [G. Gour and G. Yu, Quantum 2, 81 (2018).]. 
\end{adjustwidth*}

\justify
\begin{center}
\Large
\section{Introduction}
\end{center}

\normalsize
Discussions of quantum entanglement first came into the spot light in 1935 [1], the phenomenon did not have its current name and was instead called by Einstein-Podolsky-Rosen (EPR) as "spooky action at a distance". This "spooky action" can be characterized through the following example. Suppose two particles in remote systems $A$ and $B$ are described by a joint state

\begin{equation}
\ket{\phi_1}=\ket{\psi_A}\otimes\ket{\psi_B}=\frac{(\ket{0}+\ket{1})}{\sqrt{2}}\otimes\frac{(\ket{0}+\ket{1})}{\sqrt{2}}
\end{equation}

then it is said to be separable (not entangled) because the state of particle in system $B$ is always unchanged after a quantum measurement being performed on particle in system $A$. While if the joint state is instead 

\begin{equation}
\ket{\phi_2}=\frac{\ket{0}\otimes\ket{0}+\ket{1}\otimes\ket{1}}{\sqrt{2}}
\end{equation} 

then it is called entangled because the state of the particle in system $B$ is supposedly instantaneously changed after a measurement in remote system $A$. 

However, the previous examples of states $\ket{\phi_1}$ and $\ket{\phi_2}$ only represent respectively the states that are not entangled and those that are maximally entangled. While most states are only entangled to some degree. In the last couple of decades numerous methods of measuring the quantum entanglement have been proposed. However, it will become evident that none of them are perfect. We will examine these notions, after which we will be prepared to talk about the main subject of this paper: the monogamy of entanglement- a physical phenomenon that does not allow limitless sharing of entanglement across many subsystems. We will be reviewing the various methods of mathematical definitions of monogamous entanglement measures. 

\begin{center}
\Large
\section{Measuring entanglement}
\end{center}

\normalsize
In this section we will introduce various propositions of entanglement measures that have been developed over the years. Which is an important prerequisite to the discussion of monogamy later. 

\large
\subsection{Entanglement properties}

\normalsize

In order to create a valid measure of entanglement one must first establish properties of entanglement [2,17,25] that must be accounted for. To understand these properties, we must look at the operations that can exploit entanglement. Such operations are called Local Operations and Classical Communication (LOCC) [2]. However, it is very difficult to characterise them as there are so many different possible LOCC's. Therefore, we will instead give one of the most famous examples of LOCC that will give a general understanding of what it represents. The example is the teleportation protocol [14] which works as follows (see Figure 1). 

\begin{center}
\begin{figure}[h]
\centering
\includegraphics[scale=0.7]{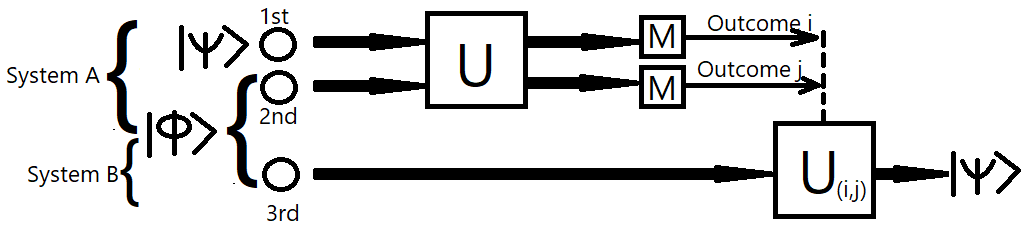}
\caption{The teleportation protocol.}
\end{figure} 
\end{center}

Suppose initially the first and the second particles are in system A and the third particle is in system B. The 1st particle is in state $\ket{\psi}$, the second and third particles make up a joint maximally entangled state $\ket{\phi}$. The joint state of all particles is $\ket{\psi}\otimes\ket{\phi}$. To follow the protocol, we begin by applying the unitary operator $U$ on particles in system A. Then the state of particles in system A is measured. Finally, depending on the outcome $(i,j)$ of the measurement a person in A classically communicates to a person in B to apply unitary $U_{(i,j)}$ on the third particle which results in it having the same state as the initial state of the first particle in A. Thus, through the teleportation protocol a person in A can send state of their particle to a person in B without sending the particle itself. 

Returning to the initially discussed subject, these are the following properties of entanglement that are to be accounted for when talking about the entanglement measures: 

\begin{adjustwidth*}{2em}{0em}

1) A separable state in a system with Hilbert space $\mathcal{H}_{A_1}\otimes...\otimes\mathcal{H}_{A_n}$ with probability distribution $p_i$ which can be written as

\begin{equation}
\rho_{A_1...A_n}=\sum_i{p_i(\rho^{A_1}_i\otimes...\otimes\rho^{A_n}_i)}
\end{equation}

has no entanglement. 

2) Entanglement does not increase under LOCC on average. 

So if we have LOCC operation where we have initial state $\rho$ transform into $\rho_i$ with probability $p_i$ then $E(\rho)\geq\sum_i{p_i E(\rho_i)}$, where $E$ is an entanglement measure. 

3) The maximally entangled state of $\rho_{AB}$ in a composite system with Hilbert space $\mathcal{H}_A\otimes\mathcal{H}_B$ is given by the pure state

\begin{equation}
\frac{\ket{0}\otimes\ket{0}+...+\ket{d-1}\otimes\ket{d-1}}{\sqrt{d}}
\end{equation}

if each partial state of the systems A and B is in d-dimension.  

4) Additivity: $E(\rho^{\otimes n})=n E(\rho)$ for all $n\in \mathbb{N}$.

5) Convexity: $E(\sum_i{p_i\rho_i}) \leq \sum_i{p_i}E(\rho_i)$. 

6) Continuity: $E(\rho)-E(\sigma)\rightarrow 0$ as $\abs{\abs{\rho-\sigma}}\rightarrow 0$.

\end{adjustwidth*}

\subsection{Entanglement measure of pure states}

The universally accepted entanglement measure of pure states is the Von Neumann entropy of entanglement, which was introduced in [13] and then was further proven to be a good measure in [3]. Suppose we want to measure entanglement between the particle(s) in system A and the particle(s) in system B that share a joint pure state $\ket{\psi}$ of the composite system with Hilbert space $\mathcal{H}_A\otimes\mathcal{H}_B$, then the Von Neumann entropy of entanglement in this case is given by

\begin{equation}
S(\ket{\psi})=-Tr(\rho_A \log_2(\rho_A))=-Tr(\rho_B \log_2(\rho_B))
\end{equation}

where $\rho_A=Tr_B(\ket{\psi}\bra{\psi})$ and $\rho_B=Tr_A(\ket{\psi}\bra{\psi})$. This is because this entanglement measure perfectly satisfies all of the properties that we introduced. Properties 1, 3, 4 can be easily checked, while properties 2, 5 and 6 were proven in [3,27,28].

\large
\subsection{Entanglement measures of mixed states}

\normalsize

However, in a laboratory we would mostly be dealing with mixed states, not the pure states. Therefore, one would be most interested in utilising a method for measuring entanglement of mixed states. Fortunately, there have been made numerous propositions of measures of entanglement of mixed states. The following most renowned propositions are set to measure the entanglement between the particle(s) in system A and the particle(s) in system B that share a joint state $\rho$ of the composite system with Hilbert space $\mathcal{H}_A\otimes\mathcal{H}_B$:

\begin{adjustwidth*}{0em}{2em}
1) Entanglement of formation [11].

\begin{equation}
E_F(\rho)=\min_{\{p_i,\ket{\phi_i}\}}\left(p_i\sum_i{S(\ket{\phi_i})}\right)
\end{equation}

where $\rho=\sum_i{p_i\ket{\phi}}$.

2) Relative entropy of entanglement [4].

\begin{equation}
E_R(\rho)=\inf_{\sigma}\{Tr(\rho log_2(\rho)-\rho log_2(\sigma))\}
\end{equation}

where $\sigma$ stands for all separable states. 

3) Squashed entanglement [5].

\begin{equation}
E_S(\rho=\rho_{AB})=\inf\{I(\rho_{ABE}/2 : Tr_E(\rho_{ABE})=\rho_{AB}\}
\end{equation}

where $I(\rho_{ABE})=S(\rho_{AE})+S(\rho_{BE})-S(\rho_{ABE})-S(\rho_{E})$.

4) Entanglement cost [6].

\begin{equation}
E_C(\rho)=\inf\{r : \lim_{n\rightarrow \infty}(\inf_{\psi}(Tr\abs{\rho^{\otimes n}-\psi(\phi(2^{rn}))}))=0\}
\end{equation}

where $\psi$ stands for a trace preserving LOCC operation and $\phi(K)$ is a maximally entangled state in $K$ dimensions. 

5) Distillable Entanglement [11].

\begin{equation}
E_D(\rho)=\sup\{r : \lim_{n\rightarrow \infty}(\inf_{\psi}(Tr\abs{\psi(\rho^{\otimes n})-\phi(2^{rn})}))=0\}
\end{equation}

where we have the same symbol meaning as in entanglement cost. 

\end{adjustwidth*}

But unfortunately, to this day we still don't know of the perfect entanglement measure. Indeed, each of the above proposals are known to have flaws.

\begin{adjustwidth*}{2em}{0em}

$\bullet$ Relative entropy of entanglement has been proven to be non-additive [7] (chapter V, subsection B). 

$\bullet$ Squashed entanglement, whose name originates from the fact that $E_D<E_S<E_F$ (proven in [5]), actually has been proven to satisfy all of the entanglement properties: continuity (proven in [8]), additivity, convexity, vanishes for separable states, non-increasing under LOCC (all four proven in [5]). However, there has not been found an easy way to determine the value of squashed entanglement for an arbitrary state.

$\bullet$ Entanglement cost, just like the squashed entanglement, also perfectly satisfies all of the properties because it has been proven in [6] that $E_C(\rho)=\lim_{n\rightarrow \infty}(E_F(\rho^{\otimes n})/n)$, which will be understood once we will go over the properties that entanglement of formation $E_F$ satisfies in the next subchapter. However, again just like the squashed entanglement, it is too difficult to compute. 

$\bullet$ Distillable entanglement is not only difficult to compute like the above two entanglements, but there is also evidence that it is neither additive nor convex [9].

$\bullet$ Lastly there is the entanglement of formation, but it is worth more than just a short mention, therefore we will be discussing it more thoroughly in the next subsection. 

\end{adjustwidth*}

\large
\subsection{Entanglement of formation}

\normalsize
Entanglement of formation is the most well-known proposal of entanglement measure of mixed states. It used to give high hopes of being the ideal entanglement measure until when it was proven to be non-additive [10]. It also makes sense to talk about entanglement of formation in greater detail because we will use some elements from this subchapter later, when we will be talking about monogamy of entanglement. 

It is not easy to derive its value with the formula that we have at the moment. Fortunately, over the course of mainly three papers [3,29,30], the entanglement of formation of an arbitrary state $\rho$, describing entanglement between two qubits, has been proven to be equal to 

\begin{equation}
E_F(\rho)=-\frac{1+\sqrt{1-C^2}}{2}\log_2\left(\frac{1+\sqrt{1-C^2}}{2}\right)-\frac{1-\sqrt{1-C^2}}{2}\log_2\left(\frac{1-\sqrt{1-C^2}}{2}\right)
\end{equation}

where $C=\max\{0, \lambda_1-\lambda_2-\lambda_3-\lambda_4\}$ (concurrence) with $\lambda_i$ being the eigenvalues of $R=\sqrt{\rho\tilde{\rho}}$ in descending order as $i$ increases, where $\tilde{\rho}=(\sigma_y\otimes\sigma_y)\rho^*(\sigma_y\otimes\sigma_y)$ with $\sigma_y$ being the y-Pauli matrix. 

\begin{center}
\Large
\section{Monogamy of entanglement}
\end{center}

\normalsize

Monogamy of entanglement [15] can be best described as a physical phenomenon that does not allow unlimited distribution of entanglement across many subsystems. It has been mathematically proven to be a valid property of entanglement in [18]. 

So why have we not put this physical phenomenon in the subchapter 3.1 as the seventh property of the entanglement? The reason for this is that there still does not exist a single universally agreed mathematical assessment of weather an entanglement measure is monogamous or not. 

The best example demonstrating monogamy is the following. Suppose we have three particles in a tripartite pure state in a composite system with Hilbert space $\mathcal{H}_A\otimes\mathcal{H}_B\otimes\mathcal{H}_C$. Each particle is in their respective two-dimensional partial state. And suppose that the first and the second particles are maximally entangled. Then monogamy manifests itself by not allowing any entanglement shared between the third particle and the other two. This can be checked if we assume entanglement between all 3 particles, to then get a contradiction, with a use of Schmidt decomposition. 

However, this is one of the extreme examples. If the first and the second particles are only entangled to some degree, then some entanglement can be shared with the third one. So, the main question is: how do we describe this mathematically? 

\large
\subsection{Monogamy of concurrence}

\normalsize

One of the first times the notion of monogamy was captured mathematically in [16]. And it goes as follows. 

\begin{theorem}
For any pure joint sate $\ket{\psi}$ of three qubits in a composite system with Hilbert space $\mathcal{H}_A\otimes\mathcal{H}_B\otimes\mathcal{H}_C$ this inequality is true

\begin{equation}
C^2_{AB}(\rho_{AB})+C^2_{AC}(\rho_{AB})\leq C^2_{A(BC)}(\rho_{ABC})
\end{equation}
\end{theorem}

Here $C_{AB}$ stands for concurrence corresponding to the entanglement of formation (see formula (11)) that measures entanglement between the particles in systems A and B, $C_{AC}$ is analogous and $C_{A(BC)}$ corresponds to entanglement between the particles in system A and the composite system of the systems B and C with Hilbert space $\mathcal{H}_A\otimes\mathcal{H}_B$.

To understand the meaning behind this inequality we must first talk about the concurrence itself. First of all, we must mention that squared concurrence has as much right to be used as a measure of entanglement between two qubits in a joint mixed state as the entanglement of formation has. The reason for this is given by the derived formula of entanglement of formation between qubits

\begin{equation}
E_F(\rho)=-\frac{1+\sqrt{1-C^2}}{2}\log_2\left(\frac{1+\sqrt{1-C^2}}{2}\right)-\frac{1-\sqrt{1-C^2}}{2}\log_2\left(\frac{1-\sqrt{1-C^2}}{2}\right)
\end{equation}

in terms of concurrence $C$. Let us plot the value of the entanglement of formation against the squared concurrence (see Figure 2).

\begin{center}
\begin{figure}[h]
\centering
\includegraphics[scale=0.8]{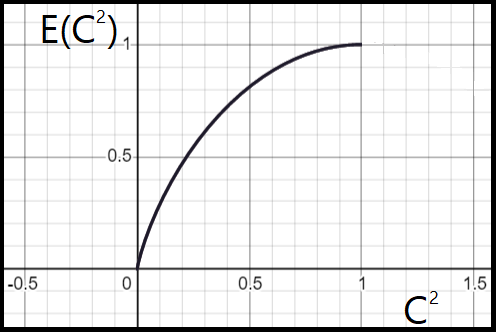}
\caption{The plot of $E_F$ against $C^2$.}
\end{figure} 
\end{center}

As we can see form the plot, the entanglement of formation is a monotonically increasing function of squared concurrence, both are equal to zero for separable states and both are equal to one for maximally entangled states. And it follows all of the other entanglement properties that entanglement of formation follows (like convexity, which has been proven to be satisfied by concurrence in [12]). This confirms the claim about the square concurrence as an entanglement measure of two qubits. 

Now let us return to the inequality (12) to see exactly how it describes monogamy. $C^2_{AB}$ and $C^2_{AC}$ describe the amount of entanglement between two particles in their respective systems. $C^2_{A(BC)}$ describes the amount of entanglement between the particle in A and the two particles in a joint state of composite system, which can be considered as a single particle in a two dimensional state. So, the inequality (12) means that the amount of entanglement between the particle in A and the other two particles bounds the amount of entanglement between the particles in systems A and B plus the amount of entanglement between the particles in A and C. So now we see how the inequality (12) describes limitations on the amount of entanglement being distributed across a tripartite system, which is a clear demonstration of monogamy. Let us plot $C^2_{AB}$ against $C^2_{AC}$ to visualise this better (see Figure 3).

\begin{center}
\begin{figure}[h]
\centering
\includegraphics[scale=0.5]{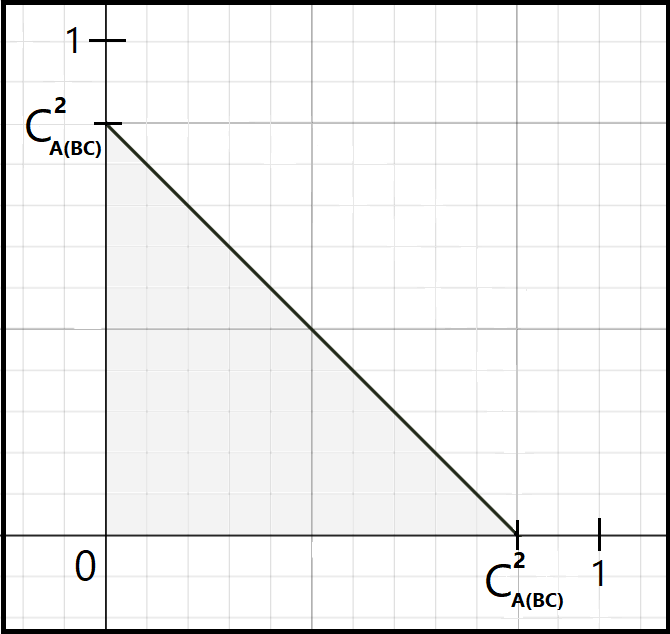}
\caption{The plot of $C^2_{AB}$ against $C^2_{AC}$.}
\end{figure} 
\end{center}

When plotting the inequality (12) one must remember that the values of squared concurrences can be only found between zero and one. Therefore, we end up with a triangular region. If we look at the plot, then we can see that monogamy manifests itself in the following way. The values of ($C^2_{AB}$, $C^2_{AC}$) can never be found outside of the triangle, which is defined by the axis intersection values $C^2_{A(BC)}$, they are constrained inside of it. 

The inequality (12) does not work for all tripartite mixed states $\rho$, however. This is because we cannot always treat state $\rho_{BC}$ as a two-dimensional state since it does not always have at least two non-zero eigenvalues, which implies that we cannot use the entanglement of formation formula (13) since it only works for measuring the entanglement of a joint state of two qubits. And therefore $C^2_{A(BC)}$ is undefined.  

But we can come up with a similar inequality, which is true for all tripartite mixed states $\rho$ of three qubits in a composite system with Hilbert space $\mathcal{H}_A\otimes\mathcal{H}_B\otimes\mathcal{H}_C$ [16]

\begin{equation}
C^2_{AB}+C^2_{AC}\leq \min_{p_i,\ket{\psi_i}}\left(\sum_i p_iC^2_{A(BC)}(\ket{\psi_i}\bra{\psi_i})\right)
\end{equation}

where $\rho=\sum_i p_i\ket{\psi_i}\bra{\psi_i}$. 

This inequality also displays monogamous properties of squared concurrence. Just like the inequality (12) for pure states it restricts the possible values of $(C^2_{AB},C^2_{AC})$ for each tripartite state $\rho_{ABC}$. 

So far, we have talked about mathematical expressions showing the monogamy of concurrence only. But what about other entanglement measures? Suppose again we have pure tripartite state of three qubits in a system with Hilbert space $\mathcal{H}_A\otimes\mathcal{H}_B\otimes\mathcal{H}_C$. Then it can be shown with a simple example that inequality

\begin{equation}
E_{AB}+E_{AC}\leq E_{A(BC)}
\end{equation}

does not work for the entanglement of formation. If we have a pure tripartite state 

\begin{equation}
\frac{1}{\sqrt{2}}(\ket{100}+\frac{1}{\sqrt{2}}\ket{010}+\frac{1}{\sqrt{2}}\ket{001})
\end{equation}

then we get $E_{AB}\approx0.6$, $E_{AC}\approx0.6$ and $E_{A(BC)}=1$ which contradicts the inequality (12). But that does not really mean that entanglement of formation does not follow the monogamy property. Indeed, the inequality (12) being true for all tripartite pure states of three qubits is a legitimate mathematical characterisation which shows that concurrence is a monogamous measure of entanglement. The same inequality has even been proven (in [19], chapter III, page 4) to be valid for the squashed entanglement and the one-way distillable entanglement for all dimensions of the tripartite state $\rho$ (the one-way distillable entanglement is the type of distillable entanglement where classical communications can only go one way: from system A to system B [20]). But that does not mean that this must be the test of monogamy for all other entanglement measures as well. Because there are indeed other entanglement measures that do not follow the inequality (12), just like the entanglement of formation or the distillable entanglement. But what if these entanglement measures could follow some different inequality for all tripartite states that could indicate monogamous properties for them? 

\large
\subsection{Monogamy of other entanglement measures}

\normalsize
Following the ending of the previous subchapter we would like to have a singular inequality for any entanglement measure which is a satisfying test of weather that entanglement measure is monogamous or not. We will see a similar approach as in the inequality (12), but much more flexible [21]. And we will see whether our new inequality will indicate monogamous properties of the entanglement measures like the entanglement of formation and the relative entropy of entanglement. 

\begin{definition} If there exists a function $f:R\times R\rightarrow R$ (where $R$ is the real numbers in $[0,\infty)$ interval) such that 

\begin{equation}
E_{A(BC)}(\rho_{ABC})\geq f(E_{AB}(\rho_{AB}),E_{AC}({\rho_{AC}}))
\end{equation}

is true for all tripartite states $\rho_{ABC}$ in a composite system with Hilbert space $\mathcal{H}_A\otimes\mathcal{H}_B\otimes\mathcal{H}_C$, then the entanglement measure $E$ is monogamous. 
\end{definition}

However, right now this statement as a verification of monogamy is much broader than what we should have been aiming for. The reason for this is that just local operations [17] (without classical communication) do not increase entanglement. Therefore, entanglement must not increase under partial trace. So, this means that 

\begin{equation}
E_{A(BC)}(\rho_{ABC})\geq E_{AB}(\rho_{AB}) \,\,\,\,\, and \,\,\,\,\, E_{A(BC)}(\rho_{ABC})\geq E_{AC}(\rho_{AC})
\end{equation}

by default. This means that with our current test of whether an entanglement measure is monogamous we can just pick $f(E_{AB}(\rho_{AB}),E_{AC}({\rho_{AC}}))=\max(E_{AB}(\rho_{AB}),E_{AC}(\rho_{AC}))$ to satisfy the inequality (18) and, therefore, make any entanglement measure monogamous automatically, making our definition useless. Therefore, since

\begin{equation}
E_{A(BC)}(\rho_{ABC})\geq \max(E_{AB}(\rho_{AB}),E_{AC}(\rho_{AC}))
\end{equation}

we should restrict our inequality inside the Definition 1 to instead be

\begin{equation}
E_{A(BC)}(\rho_{ABC})\geq f(E_{AB}(\rho_{AB}),E_{AC}({\rho_{AC}}))>\max(E_{AB}(\rho_{AB}),E_{AC}(\rho_{AC}))
\end{equation}

Though a small correction should be made to this inequality. We should be allowed to have $f=\max(E_{AB},E_{AC})$ for some values of $E_{AB}$ and $E_{AC}$. We just need to make sure there exists a two-dimensional area of $(E_{AB},E_{AC})$ for which we strictly have $f(E_{AB},E_{AC})>\max(E_{AB},E_{AC})$ (this way we still don't get every entanglement measure to be monogamous by default by our new definition). Thus, let us instead have the following definition:

\begin{definition} If there exists a function $f:R\times R\rightarrow R$ (where $R$ is the real numbers in $[0,\infty)$ interval) such that 

\begin{equation}
E_{A(BC)}(\rho_{ABC})\geq f(E_{AB}(\rho_{AB}),E_{AC}({\rho_{AC}}))\gtrdot\max(E_{AB}(\rho_{AB}),E_{AC}(\rho_{AC}))
\end{equation}

is true for all tripartite states $\rho_{ABC}$ in a composite system with Hilbert space $\mathcal{H}_A\otimes\mathcal{H}_B\otimes\mathcal{H}_C$, then the entanglement measure $E$ is monogamous. Where the meaning behind $\gtrdot$ is that there can be equality for some values of $E_{AB}$ and $E_{AC}$, but there must be a space of $(E_{AB},E_{AC})$ for which $f(E_{AB},E_{AC})>\max(E_{AB},E_{AC})$
\end{definition}

\begin{center}
\begin{figure}[h]
\centering
\includegraphics[scale=0.5]{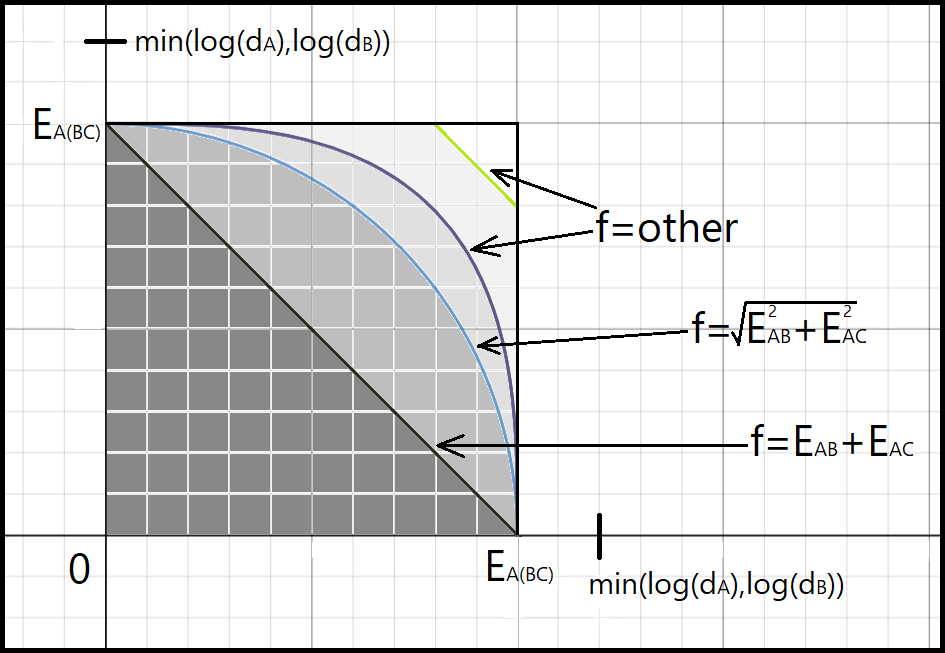}
\caption{The possible plots of $E_{AB}$ against $E_{AC}$.}
\end{figure} 
\end{center}

\newpage
For the visualisation of the Definition 2 let us give some examples (see Figure 4) of the functions $f(E_{AB},E_{AC})$ that would satisfy the inequality (21). For example, $f=E_{AB}+E_{AC}$ meets the requirements. We have $f=E_{AB}+E_{AC}=\max(E_{AB},E_{AC})$ only for $(0,E_{A(BC)})$ and $(E_{A(BC)},0)$. This is an acceptable correction because for the rest of the space of $(E_{AB},E_{AC})$ we have $f(E_{AB},E_{AC})>\max(E_{AB},E_{AC})$, as required. The function $f=\sqrt{E_{AB}^2+E_{AC}^2}$ also satisfies the inequality in an identical way to the previous one. And then there are other functions. For example, even a function defined like

\begin{equation}
f=\begin{cases}
      \max(E_{AB},E_{AC}) & \text{for}\,\,\, E_{AB},E_{AC}\in[0,\frac{4}{5}E_{A(BC)})\\ 
      E_{AB}+E_{AC}-\frac{4}{5}E_{A(BC)} & \text{for}\,\,\, E_{AB},E_{AC}\in(\frac{4}{5}E_{A(BC)},E_{A(BC)}]
    \end{cases}
\end{equation}

also satisfies the inequality (21). This is because, again, there exists a space of $(E_{AB},E_{AC})$ such that $f(E_{AB},E_{AC})>\max(E_{AB},E_{AC})$, namely the space $E_{AB},E_{AC}\in(\frac{4}{5}E_{A(BC)},E_{A(BC)}]$. For this function, the inequality (21) generates area filling most of the square except for the small corner at the top right, as seen in the Figure 4.

However, it turns out that the entanglement of formation and relative entropy of entanglement are not monogamous under the newest broader definition [21,22]. Even their regularised versions, $\lim_{n\rightarrow \infty}(E_F(\rho^{\otimes n})/n)$ and $\lim_{n\rightarrow \infty}(E_R(\rho^{\otimes n})/n)$, do not follow the new definition of monogamous entanglement measures either [21,22]. These regularised entanglement measures were considered to be perfect, following every mentioned entanglement property except for monogamy, even in the sense of our very flexible inequality. 

But perhaps our new definition of monogamous entanglement measures is actually a bit too demanding. If we alter it, so that we check if there exists a function $f$ for every fixed dimension of $\mathcal{H}_A\otimes\mathcal{H}_B\otimes\mathcal{H}_C$ and all of the tripartite states $\rho_{ABC}$ of the system with that Hilbert space (we don't consider infinite dimensions), then it can be shown that the entanglement of formation and the regularised relative entropy of entanglement are monogamous in the sense of this newest definition. Specifically, for every tripartite state $\rho_{ABC}$ in a composite system with Hilbert space $\mathcal{H}_A\otimes\mathcal{H}_B\otimes\mathcal{H}_C$ it was found that for the entanglement of formation we have

\begin{equation}
\begin{split}
E^F_{A(BC)}(\rho_{ABC})\geq \max(&E^F_{AB}(\rho_{AB})+\frac{c}{d_Ad_C\log_2(\min(d_A,d_C))^8}E^F_{AC}(\rho_{AC})^8,\\
&E^F_{AC}(\rho_{AC})+\frac{c}{d_Ad_B\log_2(\min(d_A,d_B))^8}E^F_{AB}(\rho_{AB})^8)>\\
&\gtrdot\max(E^F_{AB}(\rho_{AB}),E^F_{AC}(\rho_{AC}))
\end{split}
\end{equation}

and for the regularised relative entropy of entanglement, we have 

\begin{equation}
\begin{split}
E^{R\infty}_{A(BC)}(\rho_{ABC})\geq \max(&E^{R\infty}_{AB}(\rho_{AB})+\frac{c}{d_Ad_C\log_2(\min(d_A,d_C))^4}E^{R\infty}_{AC}(\rho_{AC})^4,\\
&E^{R\infty}_{AC}(\rho_{AC})+\frac{c}{d_Ad_B\log_2(\min(d_A,d_B))^4}E^{R\infty}_{AB}(\rho_{AB})^4)>\\
&\gtrdot\max(E^{R\infty}_{AB}(\rho_{AB}),E^{R\infty}_{AC}(\rho_{AC}))
\end{split}
\end{equation}

where $d_A$, $d_B$ and $d_C$ are the dimensions of the partial states $\rho_A$, $\rho_B$ and $\rho_C$ respectively [21,22]. 

Thus, we see that indeed for fixed dimensions the values of $(E_{AB},E_{AC})$ are bounded by a function $f$ defined by (21), and therefore, these entanglement measures display monogamous properties in terms of the Definition  2 with fixed dimensions. Because we have shown that indeed there is an area of values $(E_{AB},E_{AC})$ that can't exist inside the square defined by $E_{A(BC)}\geq\max(E_{AB},E_{AC})$, meaning that monogamy can't be shared limitlessly. 

To visualise this, we present plots of $E_{AB}$ against $E_{AC}$ for the respective entanglement measures in the case of a tripartite state of tree qubits in the Figure 5.

\begin{figure}[h]
\centering
\begin{subfigure}{.5\textwidth}
  \centering
  \includegraphics[width=0.8\linewidth]{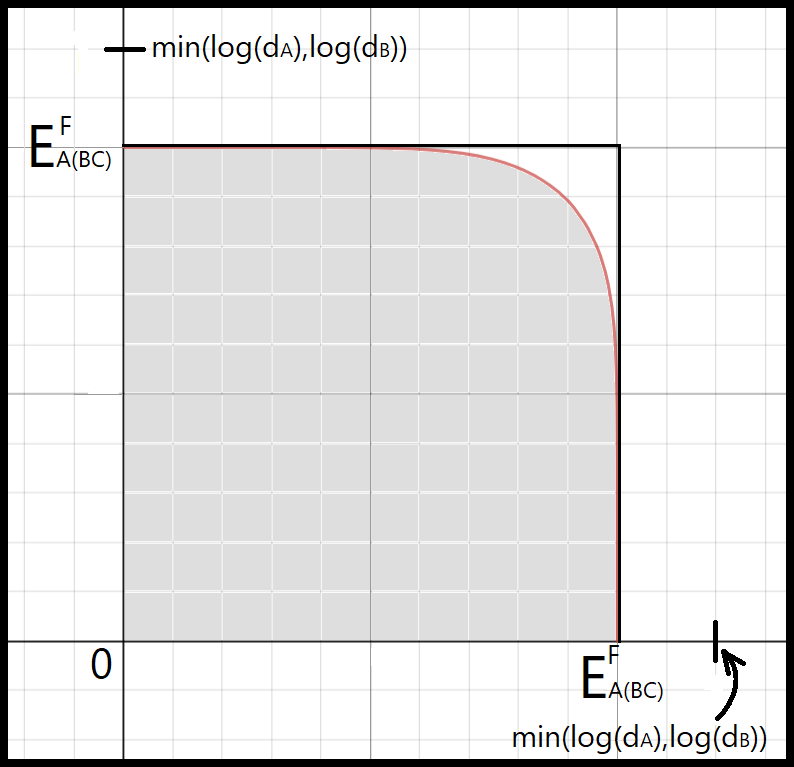}
  \caption{Entanglement of formation.}
  \label{fig:sub1}
\end{subfigure}%
\begin{subfigure}{.5\textwidth}
  \centering
  \includegraphics[width=0.8\linewidth]{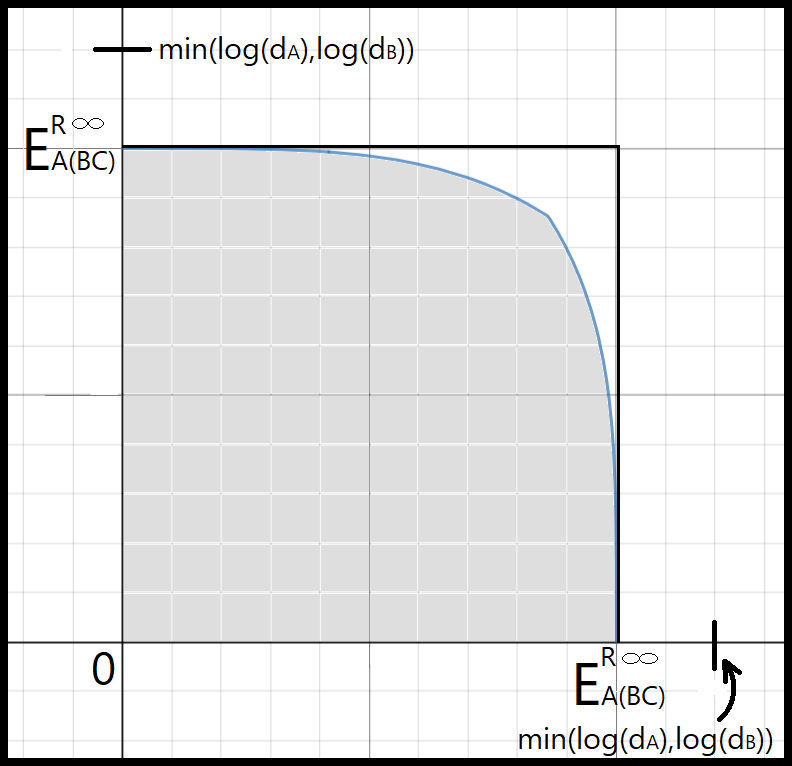}
  \caption{Regularised entropy of entanglement.}
  \label{fig:sub2}
\end{subfigure}
\caption{The two plots of $E_{AB}$ against $E_{AC}$ with $d_A=d_B=d_C=2$.}
\label{fig:test}
\end{figure}

Thus, the entanglement of formation remains to be imperfect even though it was confirmed that it is monogamous. However, we know that the regularised relative entropy of entanglement follows all of the properties of entanglement, because it has been shown that for finite dimensional states entanglement can't be shared limitlessly for this measure. This means that it is in the same group as the squashed entanglement- perfectly satisfying all of the properties, but still very difficult to compute. 

\large
\subsection{Monogamy in terms of equalities?}

\normalsize
The paper [23] argues that the definition of monogamous entanglement measures given in [21] is not the best one. We will give the definition of monogamous measurement that was introduced in [23] and then compare it to the previous definition from [21].

\begin{definition}
An entanglement measure $E$ is monogamous if for all tripartite states $\rho_{ABC}$ of composite system with Hilbert space $\mathcal{H}_A\otimes\mathcal{H}_B\otimes\mathcal{H}_C$ such that

\begin{equation}
E_{A(BC)}(\rho_{ABC})=E_{AB}(\rho_{AB})
\end{equation}

we have $E_{AC}(\rho_{AC})=0$
\end{definition}

So visually the Definition 3 means that an entanglement measure $E$ is monogamous if the values of $(E_{AB}(\rho_{AB}),E_{AC}(\rho_{AC}))$ can only be found on the interior of the square generated by 

\begin{equation}
E_{A(BC)}(\rho_{ABC})\geq\max(E_{AB}(\rho_{AB}),E_{AC}(\rho_{AC}))
\end{equation}

and on the two bottom left edges of that square (see Figure 7). Since the values of $(E_{AB}(\rho_{AB}),E_{AC}(\rho_{AC}))$ can never be found outside of the square generated by (26). Because entanglement is non-increasing under local operations, specifically the partial tracing. 

\begin{center}
\begin{figure}[h]
\centering
\includegraphics[scale=0.35]{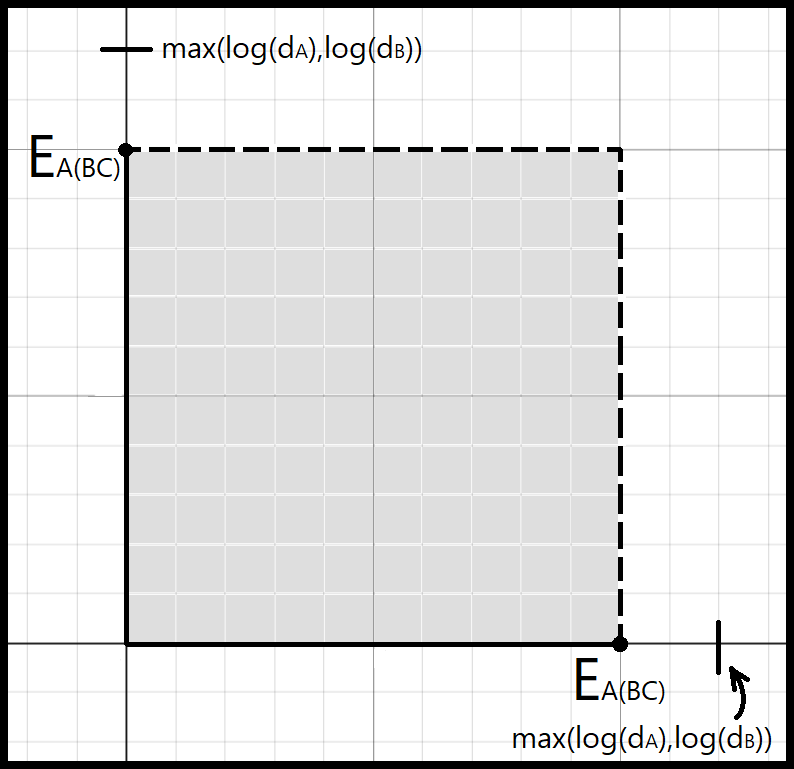}
\caption{The plot of $E_{AB}$ against $E_{AC}$.}
\end{figure} 
\end{center}

One can confirm that with this newest definition of monogamy, entanglement measures such as: squared concurrence (only for 3 qubits), entanglement of formation and regularised relative entropy of entanglement are monogamous. For the squared concurrence one just needs to insert $C^2_{A(BC)}(\rho_{ABC})=C^2_{AB}(\rho_{AB})$ into the inequality (12) to see that the only possible value of $E_{AC}(\rho_{AC})$ is zero. The proof that entanglement of formation and regularised relative entropy are monogamous in terms of the newest definition will automatically follow after Proof 1 below. 

But before that let us begin with noting that the newest definition of monogamous entanglement measures does not agree with the definition from [21], which was introduced in the previous subchapter. This is evident because the Definition 2 allows  $E_{AC}(\rho_{AC})$ to be non-zero when $E_{A(BC)}(\rho_{ABC})=E_{AB}(\rho_{AB})$. The best example is the function $f$ given by (22), which allows all of the values of $(E_{AB}(\rho_{AB}),E_{AC}(\rho_{AC}))$ inside the square generated by (26), except for the small top right corner of the square (see Figure 4). 

But the Definition 2 of monogamous entanglement measures becomes equivalent to the Definition 3, if we revise it as follows (the proof will be given in the Proof 1 below).

\begin{definition}
If there exists a function $f:R\times R\rightarrow R$ (where $R$ is the real numbers in $[0,\infty)$ interval) such that 

\begin{equation}
E_{A(BC)}(\rho_{ABC})\geq f(E_{AB}(\rho_{AB}),E_{AC}({\rho_{AC}}))\geq\max(E_{AB}(\rho_{AB}),E_{AC}({\rho_{AC}}))
\end{equation}

is true for every fixed dimension $d=dim(\mathcal{H}_A\otimes\mathcal{H}_B\otimes\mathcal{H}_C)$ and all tripartite states $\rho_{ABC}$ in a composite system with Hilbert space $\mathcal{H}_A\otimes\mathcal{H}_B\otimes\mathcal{H}_C$, and such that 

\begin{equation}
f(E_{AB}(\rho_{AB}),E_{AC}({\rho_{AC}}))=\max(E_{AB}(\rho_{AB}),E_{AC}(\rho_{AC}))
\end{equation}

is only true at $(E_{AB}(\rho_{AB}),E_{AC}(\rho_{AC}))=(0,E_{A(BC)}(\rho_{ABC}))$ and $(E_{AB}(\rho_{AB}),E_{AC}(\rho_{AC}))=$ \newline$=(E_{A(BC)}(\rho_{ABC}),0)$, then the entanglement measure $E$ is monogamous. 
\end{definition}

Visually (see Figure 7) this definition means that an entanglement measure $E$ is monogamous if for all $\rho_{ABC}$ the values of $(E_{AB}(\rho_{AB}),E_{AC}(\rho_{AC}))$ can only be found inside the area confined by the two axes $E_{AB}(\rho_{AB})$ and $E_{AC}(\rho_{AC})$, and some curve $E_{A(BC)}(\rho_{ABC})=f(E_{AB}(\rho_{AB}),E_{AC}({\rho_{AC}}))$, such that it joins together the points $(E_{A(BC)}(\rho_{ABC}),0)$ and $(0,E_{A(BC)}(\rho_{ABC}))$, and such that it is entirely inside (except for its two endpoints) the interior of the square area generated by (26). 

\begin{center}
\begin{figure}[h]
\centering
\includegraphics[scale=0.5]{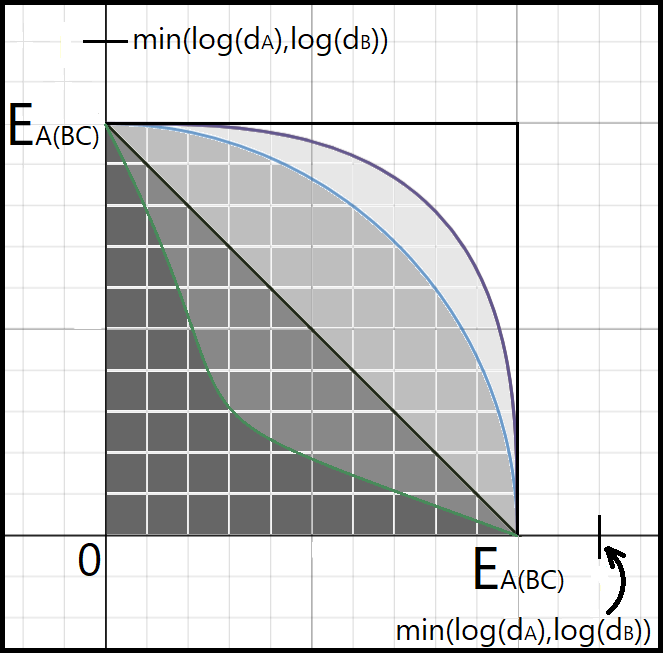}
\caption{The plot of $E_{AB}$ against $E_{AC}$.}
\end{figure} 
\end{center}

\newpage
\begin{proof} 

To prove that definitions 3 and 4 are equivalent we will start with a theorem [21]. 

\begin{theorem}
An entanglement measure $E$ is monogamous in the sense of the Definition 3 if and only if there exists $0<\alpha<\infty$ for every fixed dimension $d=dim(\mathcal{H}_A\otimes\mathcal{H}_B\otimes\mathcal{H}_C)$ such that 

\begin{equation}
E_{A(BC)}(\rho_{ABC})\geq (E_{AB}(\rho_{AB})^{\alpha}+E_{AC}(\rho_{AC})^{\alpha})^{1/\alpha}
\end{equation}

for all tripartite states $\rho_{ABC}$ of a composite system with Hilbert space $\mathcal{H}_A\otimes\mathcal{H}_B\otimes\mathcal{H}_C$. 
\end{theorem}

The inequality (29) is one of the earlier representations of monogamy. This inequality was talked about and applied to various entanglement measures in [26]. We can see what the inequality (29) looks like for various $\alpha$'s in the Figure 8. 

\begin{center}
\begin{figure}[h]
\centering
\includegraphics[scale=0.3]{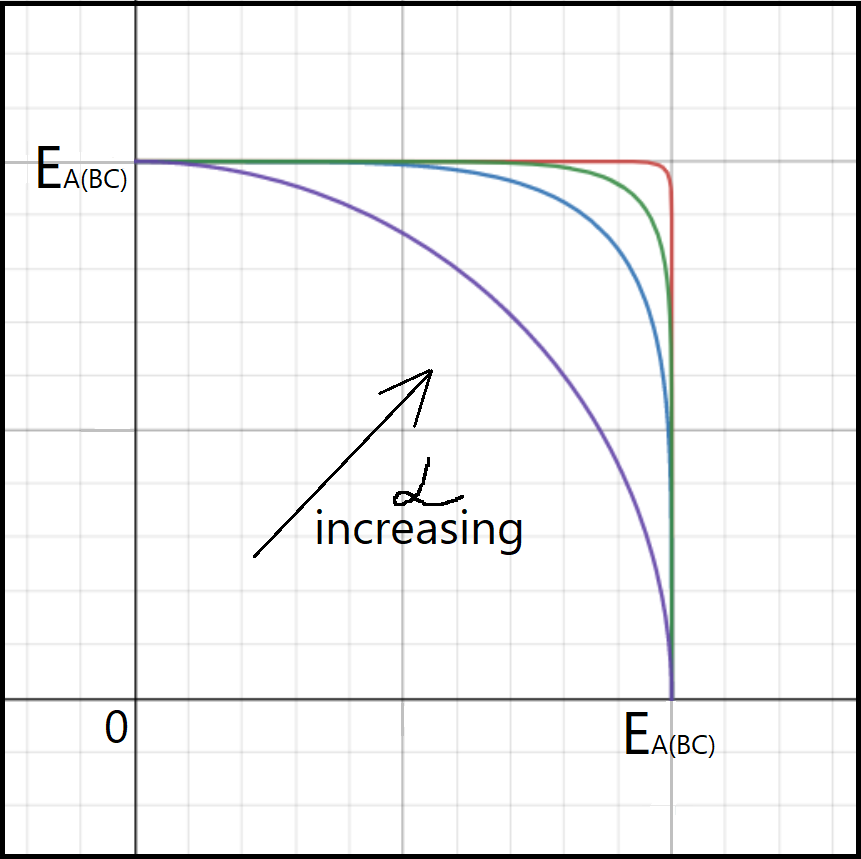}
\caption{Curves for $\alpha=2, 10, 15, 50$.}
\end{figure} 
\end{center}

Following this, the equivalence of the Definitions 3 and 4 is evident to be true because of the Theorem 2. It is easy to see that the definition of monogamy given inside the Theorem 2 is identical to the Definition 4, except that in the Definition 4 we have some function $f$, while in the Theorem 2 we have $(E_{AB}(\rho_{AB})^{\alpha}+E_{AC}(\rho_{AC})^{\alpha})^{1/\alpha}$ for some $\alpha$. Notice also that $f$ and $\alpha$ are chosen for a fixed dimension of tripartite states in both cases. While reading the proof it is advised to compare the Figures 5 and 8 for visualisation. So, to show equivalence between the Definitions 3 and 4 we need the following two-way proof:

\begin{adjustwidth*}{2em}{0em}

$(\implies)$ Suppose an entanglement measure $E$ is monogamous in the sense of the Definition 3. Then let us simply choose $f=(E_{AB}(\rho_{AB})^{\alpha}+E_{AC}(\rho_{AC})^{\alpha})^{1/\alpha}$ (by the Theorem 2). Note that we have $(E_{AB}(\rho_{AB})^{\alpha}+E_{AC}(\rho_{AC})^{\alpha})^{1/\alpha}\geq\max(E_{AB}(\rho_{AB}),E_{AC}(\rho_{AC}))$  for all $E_{AB}$, $E_{AC}\in [0,E_{A(BC)}]$ and all $\alpha>0$. Thus, the entanglement measure $E$ is monogamous in the sense of the Definition 4.

$(\impliedby)$ Suppose an entanglement measure $E$ is monogamous in the sense of the Definition 4. Then for every curve 

\begin{equation}
E_{A(BC)}(\rho_{ABC})=f(E_{AB}(\rho_{AB}),E_{AC}({\rho_{AC}}))
\end{equation}

which is entirely inside the interior of the square defined by (26), except for its two endpoints at $(E_{A(BC)}(\rho_{ABC},0)$ and $(0,E_{A(BC)}(\rho_{ABC})$, we can find an $\alpha>0$ such that 

\begin{equation}
f(E_{AB}(\rho_{AB}),E_{AC}({\rho_{AC}}))\geq (E_{AB}(\rho_{AB})^{\alpha}+E_{AC}(\rho_{AC})^{\alpha})^{1/\alpha}
\end{equation}

is true for all $E_{AB}$, $E_{AC}\in [0,E_{A(BC)}]$, therefore, the entanglement measure $E$ is monogamous in terms of the Definition 3 (by the Theorem 2).

The reason the inequality (31) can be satisfied for some $\alpha>0$ is that for every point $(E_{AB},E_{AC})$ inside the interior of the square defined by (26) we can find $\alpha>0$ such that the shape defined by inequality (29) engulfs this point. This is true because $(x^{\alpha}+y^{\alpha})^{1/\alpha}\rightarrow \max(x,y)$ as $\alpha\rightarrow\infty$. And the curve (30) is entirely made up of these interior points, except for its endpoints. Thus, we can find $\alpha>0$ such that the shape defined by (29) engulfs that curve $f$, meaning that the inequality (31) is satisfied. 

\end{adjustwidth*}

\end{proof}
\hfill $\square$

This is the original observation of this paper. We showed that the Definitions 3 and 4 are equivalent. Which automatically proves that the inequalities (23) and (24) from the previous subchapter (which were deduced in [21]) also show that the entanglement of formation and the regularised relative entropy of entanglement are monogamous in the sense of the Definition 3 (introduced in [23]). This is because these inequalities are clearly monogamous in the sense of the Definition 4. 

Authors of the paper [23], who introduced the Definition 3, seem to have made a slight oversight on this. This is because they only mentioned that the entanglement of formation has been proven to be non-monogamous in the sense of the Definition 1 (where infinite dimensions are considered), using this point as a justification that their definition of monogamous entanglement measures is better than the one from [21]. But they didn't seem to realise how close their definition of monogamous entanglement measures is to the one from [21]. Which is likely why they never talked about the inequalities (23) and (24) displaying monogamous properties of their respective entanglement measures in terms of the Definition 3. They instead came up with an entirely different proof in [24] of the fact that the entanglement of formation is monogamous in the sense of the Definition 3. When instead they could've just used (23) and (24) to prove that entanglement of formation AND regularised relative entropy of entanglement are monogamous in the sense of their definition.

\begin{center}
\Large
\section{Conclusion}
\end{center}

\normalsize
In this paper we have had a brief and informative introduction to the entanglement quantification and an in-depth examination of the entanglement property, called monogamy. 

We introduced properties that must be satisfied by a "good" entanglement measures and we introduced some of the most prevalent entanglement measures that have been proposed so far. 

Finally, we were introduced to the seventh entanglement property. The property which does not allow limitless distribution of entanglement across many subsystems. Specifically, we focused on how entanglement is shared across three subsystems. 

We looked at an attempt which tried to see if the entanglement of formation and the relative entropy of entanglement are monogamous in terms of the Definition 1. It turned out that they are not. But then after revising the Definition 1 to be less demanding, it was evident that the inequalities (23) and (24) are true for the entanglement of formation and the relative entropy of entanglement, displaying monogamy in terms of the Definition 2, but for fixed dimensions only. 

Then we looked at another attempt of defining monogamous entanglement measures. This was done with the Definition 3, which used equalities instead, proposed by [23] and argued to be better than the previous definitions. However, after making comparisons of this definition with the previous one, it made us realise that actually these definitions only vary very little from each other. We mentioned that there was an entire paper dedicated to the proof that the entanglement of formation is monogamous in terms of the Definition 3. However, given that I showed that the definitions of monogamy from [21] and [23] are very similar, I was able to make an original observation and show that the inequality (23), which existed before the publication of that paper, already served as a proof of the fact that the entanglement of formation is monogamous in the sense of the Definition 3. The same observation also automatically led us to the revelation that the inequality (24) proves that the regularised relative entropy of entanglement is monogamous in the sense of the Definition 3 as well. Unlike the proof of the monogamy of the entanglement of formation, this fact has not been proven for the regularised relative entropy of entanglement before. This ultimately confirms that this entanglement measure follows all properties of entanglement perfectly.  

For the full version of this paper, which has all the proofs explained comprehensively, see [31]. 

\section{Acknowledgement}

I would like to thank Prof Gerardo Adesso, my dissertation supervisor and one of the people who worked on "Should Entanglement Measures be Monogamous or Faithful?" [21], for helping me understand/reconstruct the many proofs related to the numerous statements that were included in my dissertation. This in-depth look into the subject equipped me with knowledge to prove that inequalities (23) and (24) from his paper show that the entanglement of formation and the regularised relative entropy of entanglement are monogamous in terms of the definition utilising equalities, given in [23]. 

\bigskip

\begin{figure}[h]
\centering
\Large
\section{References}
\end{figure}

\normalsize

[1] A. Einstein, B. Podolsky, and N. Rosen, “Can quantum-mechanical
description of physical reality be considered complete?”, Phys. Rev.
47, 777 (1935). 

[2] M. Plenio and S. Virmani, Quantum Inf. Comput., 7, 1-51 (2007).

[3] C. H. Bennett, H. J. Bernstein, S. Popescu, and B. Schumacher, Phys. Rev. A 53, 2046 (1996).

[4] V. Vedral, M.B. Plenio, K. Jacobs and P.L. Knight,
Phys. Rev. A 56, 4452 (1997).

[5] M. Christandl and A. Winter, J. Math. Phys 45, 829
(2004). 

[6] P. Hayden, M. Horodecki, and B.M. Terhal, J. Phys. A
34, 6891 (2001).

[7] K. G. H. Vollbrecht, R. F. Werner, Phys. Rev. A 64, no.
6, 062307, (2001).

[8]  R. Alicki and M. Fannes, J.Phys.A:Math.Gen.37 L55 (2004).

[9] P. W. Shor, J. A. Smolin, B. M. Terhal, Phys. Rev. Lett.
86, no. 12, 2681–2684, (2001).

[10] M. B. Hastings, Nature Phys 5, 255–257 (2009).

[11] C.H. Bennett, D.P. DiVincenzo, J.A. Smolin, and W.K.
Wootters, Phys. Rev. A 54, 3824 (1996).

[12] W. K. Wootters, Phys. Rev. Lett. 80, 2245 (1998).

[13] J. Neumann, "Mathematical Foundations of Quantum Mechanics" (1932)

[14] C. Bennett, G. Brassard, C. Crépeau, R. Jozsa, A. Peres and W. Wootters, Phys. Rev. Lett. 70, 1895 (1993).

[15] B. Terhal, IBM J. Res. Dev. 48, 71 (2004).

[16] V. Coffman, J. Kundu, andW. K.Wootters, Phys. Rev. A 61,
052306 (2000).

[17] M.B. Plenio and V. Vedral, Contemp. Phys. 39, 431
(1998).

[18] D. Yang, Phys. Lett. A 360, 249 (2006).

[19] M. Koashi and A. Winter, Phys. Rev. A 69, 022309 (2004).

[20] I. Devetak and A. Winter, Proc. R. Soc. Lond. A 461,
207 (2005).

[21] C. Lancien, S. Di Martino, M. Huber, M. Piani, G. Adesso and A.Winter Phys. Rev. Lett., 117:060501 (2016).

[22] The supplementary material to [28] given at \newline http://link.aps.org/supplemental/10.1103/PhysRevLett.117.060501

[23] G. Gour and G. Yu, Quantum 2, 81 (2018).

[24] G. Gour and G. Yu, Phys. Rev. A 99, 042305 (2019).

[25] D. Bruss, J. Math. Phys. 43, 4237 (2002).

[26] Y. Luo and Y. Li. Monogamy of $\alpha$th power entanglement measurement in qubit systems. Ann. Phys., 362:511-520 (2015).

[27] A. Wehrl, Rev. Mod. Phys. 50, 221, esp. p. 237. (1978).

[28] M. Fannes, Commun. Math. Phys. 31, 291 (1973).

[29] S. Hill and W. K. Wootters, Phys. Rev. Lett. 78, 5022 (1997).

[30] W. K. Wootters, Phys. Rev. Lett. 80, 2245 (1998).

[31] Supplementary material to this paper given at https://arxiv.org/abs/2111.05109

\end{flushleft}

\end{document}